\documentclass{iopart}
\usepackage{graphicx}
\def\be{\begin{equation}}
\def\ee{\end{equation}}
\begin{document}
\jl{6}


\title{String-inspired cosmology}

\author{David Wands {\small {\sf david.wands@port.ac.uk}}}

\address{Institute of Cosmology and Gravitation, University of
  Portsmouth, Mercantile House, Portsmouth, PO1 2EG, 
  United Kingdom}



\begin{abstract}
  I discuss cosmological models either derived from, or inspired by,
  string theory or M-theory. In particular I discuss solutions in the
  low-energy effective theory and the role of the dilaton, moduli and
  antisymmetric form fields in the dimensionally reduced effective
  action. The pre big bang model is an attempt to use cosmological solutions
  to make observational predictions. I then discuss the effective
  theory of gravity found in recent brane-world models where we live
  on a 3-brane embedded in a five-dimensional spacetime and how
  the study of cosmological perturbations may enable us to test these
  ideas.\footnote{Article based on a talk presented at ``The Early
    Universe and Cosmological Observations: a Critical Review'', Cape
    Town, July 2001}
\end{abstract}


\section{Motivation}

It is not hard to think of questions in cosmology that go beyond
the framework of classical general relativity. In an expanding
universe as one traces back our past worldline to ever higher
energies and densities one reaches an initial `big bang'
singularity where the classical theory breaks down. The hope is
that a finite quantum theory of gravity exists which remains
well-defined. If string theory is the leading candidate for that
fundamental quantum theory of gravity then it should provide some
insight on the peculiar initial conditions required by the
standard big bang. If inflation is to provide a dynamical
explanation of those initial conditions~\cite{arl}, then string
theory should provide the answer to the question of what the
inflaton field is that drives it. As a higher dimensional theory,
string theory should also be able to answer why we only observe
three large spatial dimensions, and one of time. Even at low
energies there are puzzles such as the unresolved embarrassment of
the small, but apparently finite, cosmological constant (or dark
energy).

But if cosmology needs string theory, then string theory itself
needs cosmology. For a start, it has an embarrassment of riches in
that string theory (or M-theory) has a large number of possible
low-energy vacuum states.  To answer the question of why we live
in the particular physical state we observe requires a dynamical
model of how we got here. The number of large extra dimensions is
just one example of the many moduli that parameterise different
vacuum states that have to be fixed.
A more prosaic reason why I believe string theory needs cosmology
is that string theory is a grand theoretical edifice that needs
some experimental support. It may be that cosmology is the only
arena in which string theorists can make predictions that are
observable and could promote it to the status of an empirical
science.

The subject of string cosmology is far from being a mature field of
research where one could expect some consensus about what might be
considered a standard model. There are many different approaches
reflecting the many different perspectives on both string theory and
cosmology. This article inevitably reflects my own personal interests
and prejudices.

\section{What is string theory?}

At its most fundamental level, string theory is a theory for
calculating scattering amplitudes between asymptotic states where the
fundamental objects are one-dimensional strings. The spectrum of
states includes a massless spin-2 graviton and as a result string
theory has been said to `predict gravity' - though only after apples
had already been observed falling to the ground. Nonetheless gravity
is an essential component for a would-be theory of cosmology.  But a
model of asymptotic states seems rather unpromising for a framework in
which to study cosmology where we, as observers, are embedded within
the model and asymptotic states may be of limited relevance.

In practice, most calculations are done in the low-energy
limit where string theory can be described by an effective field
theory. In the mid-1980s it was shown that low-energy effective
actions can be derived for fields that are consistent low-energy
background solutions \cite{FradTsey,CGMP,Lovelace}.
All the low-energy effective actions of
string theories include a gravitational part in the action~\cite{LWC}:
\begin{equation}
\label{10DS}
 S = {1\over \ell_s^8} \int d^{10}x \sqrt{-g} e^{-\phi} \left[ R +
(\nabla\phi)^2 \ldots \right. \,,
\end{equation}
where $\ell_s$ is the fundamental string length scale, $R$ is the
Ricci curvature scalar, and $\phi$ the dilaton. The low-energy effective
theory of gravity is familiar to relativists as a Jordan-Brans-Dicke
type theory of gravity where the Brans-Dicke parameter takes the value
$\omega=-1$. If the dilaton were left free then this would be
experimentally unacceptable as a theory of gravity so the dilaton is
one of the massless fields in the low-energy effective action that
must become fixed by the present.

By the end of the 1980's string theorists had more string theories
than they knew what to do with: five low-energy effective
10-dimensional supergravity theories, to which has now been added
11-dimensional supergravity. However the string theory revolution
of the 1990s showed that by including higher-dimensional extended
objects (two-dimensional membranes or, more generally, {\em
branes}) these six-theories can be related by duality symmetries
and are now seen to be part of some more fundamental {\em
M-theory}. These dualities include T-duality which relates
solutions on a large toroidal geometry to those on a small one,
and S-duality which relates strong and weak coupling limits. While
there may be no understanding of what M-theory is (in the way that
string theory was understood as a theory of strings) it too can be studied
through its different low-energy effective descriptions.
In 11-D supergravity there is no dilaton, but there are of course
additional geometrical moduli, and a Kaluza-Klein dimensional
reduction will still give rise to a dilaton-type field describing
the size of the compact dimension.

Branes are not part of the perturbative formulation of string
theory and duality symmetries have played a key role in mapping
out non-perturbative structure of M-theory. One would certainly
hope that they could provide a valuable insight into
non-perturbative cosmology where the low-energy descriptions
breakdown.

For a more substantive review of the state of the subject see, for
example, the review by Duff~\cite{Duff}.

\section{Dilaton-gravity}

The basic building block for many string-inspired cosmologies is the
rolling radii solution of Mueller\cite{Mueller90} for the
10-dimensional dilaton-gravity action (\ref{10DS}) in a spatially-flat
and homogeneous spacetime
\[
ds^2 = -dt^2 + \sum_{i=1}^{10} a_i^2(t) dx_i^2
\]
where
\begin{equation}
\label{Mueller}
a_i\propto t^{p_i} \qquad \phi \propto K \ln t
\end{equation}
and the indices obey the generalised Kasner constraints
\[
\sum_i p_i = 1+K \,, \qquad \sum_i p_i^2 = 1 \,.
\]
I refer to these as dilaton-vacuum solutions in the sense that
only the gravitational fields (in which I include the dilaton) are
excited and no matter fields are included.
These solution are in general singular as, for example, the Ricci
curvature diverges for $K\neq0$~\cite{LWC}.
Higher-order curvature corrections are expected to become important
when the curvature approaches the string scale, $R\sim \ell^{-2}$, and
loop corrections as the dilaton approaches the strong coupling regime
$g_s^2=e^\phi\sim1$.

The original 10-D dilaton-gravity action (\ref{10DS}) can be
reduced down to 4-D if the fields (and metric) are assumed to be
independent of the six internal dimensions. Compactifying on an
internal space of volume $\propto e^{6\beta}$ would yield the
four-dimensional action
\[
S_4\ =\ {1\over 2\kappa^2}\ \int d^4 x\ \sqrt{-g}\ e^{-\varphi}\ \left[
R
 + (\nabla^2\varphi) - 6(\nabla \beta)^2 \ldots \right]
\]
where my ellipsis represent all the moduli associated with the
internal geometry other than the token modulus $\beta$ that I have
arbitrarily kept. The symmetries of the internal space determines
the physics of what can be re-interpreted as matter fields in a
four-dimensional world.

The 4D effective dilaton is given by
\[
\varphi = \phi - 6 \beta
\]
and the effective gravitational coupling $\kappa^2$ is proportional to
$\ell_s^8$ divided by the volume of the internal space.  Thus a
hierarchy may be introduced between the higher-dimensional string
scale and that apparent in the 4D theory if the internal space is much
larger than the fundamental string scale~\cite{ADD}.

Neglecting all dependence of the fields and the geometry upon the
internal space is of course a huge simplification, but it can be
done in a self-consistent manner. Arbitrary perturbations about
this background which allow inhomogeneity on the internal space
can be decomposed in eigenvalues of the Laplacian on the internal
space which then can again be treated as effective fields in the
4D world. If the compactification scale is of order the string
scale, then these are very massive excitations that are not part
of the low-energy effective theory. In theories where the scale of
compactification is much larger than the string scale it may be
necessary to include inhomogeneous bulk configurations at
intermediate energy scales.

\section{Form fields}

Much of my own work on string-inspired cosmology has investigated
the dynamical role of antisymmetric form fields that appear
alongside the dilaton and moduli in the bosonic sector of the
low-energy superstring effective actions. These have often been
ignored in previous discussions because their effective density is
typically localised in space or time and may be negligible in the
asymptotic regime.
Nonetheless they can play a crucial role interpolating between
dilaton-vacuum solutions and dynamically determining the allowed
asymptotic states \cite{CLW94,FVM}.

As an example consider a spatially flat FRW metric,
\[
ds^2 = a^2(\eta) \left[ -d\eta^2 + dx^2 + dy^2 + dz^2 \right]
\]
The model-independent axion in four-dimensions is a pseudo-scalar
$\sigma$ Poincare-dual to the anti-symmetric three-form field
strength $H=*\nabla\sigma$. The general axion-dilaton solutions
(neglecting now for simplicity the modulus field, $\dot\beta=0$)
can be given in simple closed form:
\begin{eqnarray}
\label{axidilaton} e^\varphi &=& {e^{\varphi_*} \over 2}
 \left[ \left( {\eta\over\eta_*} \right)^{-\sqrt{3}}
   + \left( {\eta\over\eta_*} \right)^{\sqrt{3}} \right]
\nonumber\\
a &=& {a_*\over\sqrt{2}} \left[ \left( \frac{\eta}{\eta_*}
\right)^{1-\sqrt{3}} +
 \left( \frac{\eta}{\eta_*} \right)^{1+\sqrt{3}} \right]^{1/2}
 \\
\sigma &=& \sigma_* \pm e^{-\phi_*} \left[
 {(\eta/\eta_*)^{-\sqrt{3}} - (\eta/\eta_*)^{\sqrt{3}} \over
   (\eta/\eta_*)^{-\sqrt{3}} + (\eta/\eta_*)^{\sqrt{3}} } \right]
\nonumber
\end{eqnarray}
These solutions provide a nice example of how the symmetries of the
string effective actions (in this case an SL(2,R) symmetry which is
related to the modular invariance of the string compactified on a
two-torus~\cite{Wilczek,LWC}) can be used to generate the general
axion-dilaton solutions starting from the dilaton-vacuum
solutions~\cite{CEW97}.

\begin{figure}[t]
\centering
\includegraphics[scale=0.8]{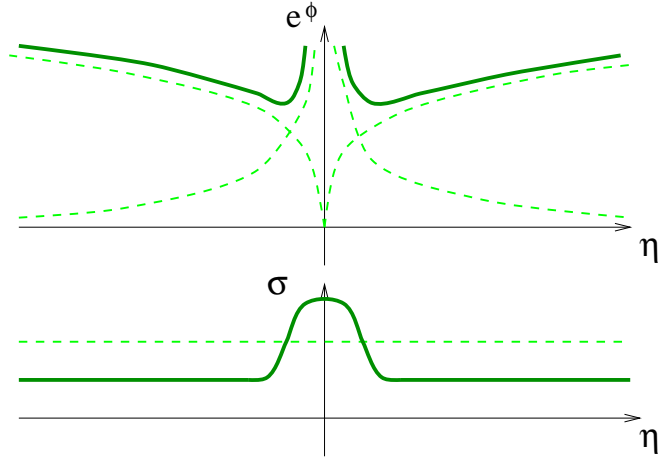}
\caption[axion]{\label{axion} Evolution of the dilaton and axion
  fields as functions of conformal time $\eta$ in
  Eq.~(\ref{axidilaton}) for a spatially flat
  Friedmann-Robertson-Walker spacetime.}
\end{figure}

The axion remains almost constant except when $\eta\sim\eta_*$
where its value undergoes a step up or down. The asymptotic
solution returns to the dilaton-vacuum solution, but the presence
of the axion nonetheless selects the allowed asymptotic solutions,
in particular placing a lower bound on the dilaton field.
Note that because the dilaton and axion are massless fields the
Einstein frame scale factor is simply $\tilde{a} =
\tilde{a}_*(\tilde{t}/\tilde{t}_*)^{1/3}$, where $\tilde{t}$ is the
proper cosmic time in the Einstein conformal frame~\cite{LWC}.

\begin{figure}[t]
\centering
\includegraphics[scale=0.8]{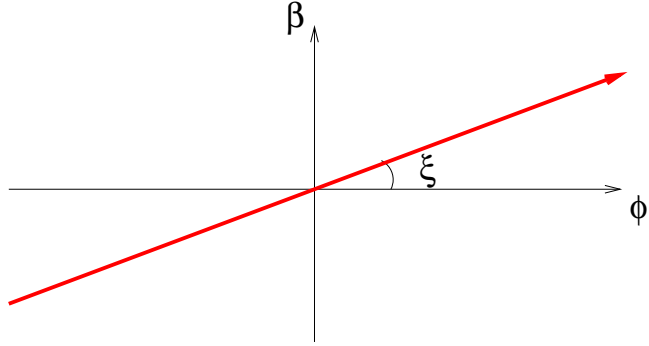}
\caption[dilmod]{\label{dilmod} Evolution of the dilaton and
  modulus fields in Mueller's rolling radii solutions.}
\end{figure}

Reintroducing the modulus $\beta$, we see that the
dilaton-moduli-vacuum solutions (\ref{Mueller}) correspond to free
particle moving in a straight line in $(\phi,\beta)$ field space
(see figure~\ref{dilmod}):
\begin{eqnarray}
\varphi = \varphi_* + \sqrt{3}(\cos\xi) \, \ln(\eta/\eta_*)  \nonumber \\
\beta = \beta_* + \sqrt{3} (\sin\xi) \, \ln(\eta/\eta_*) \nonumber
\, .
\end{eqnarray}

\begin{figure}[t]
\centering
\includegraphics[scale=0.8]{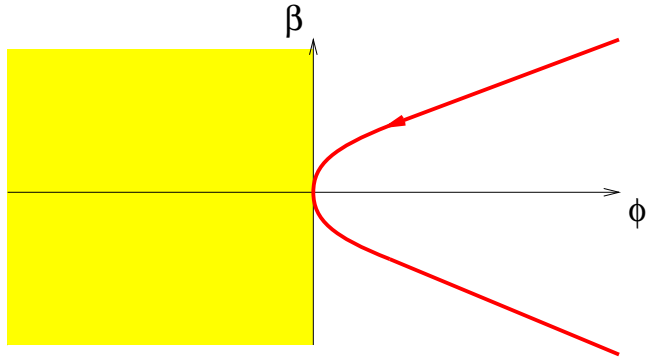}
\caption[axidilmod]{\label{axidilmod} Evolution of the dilaton and
  modulus fields in axion-dilaton-moduli solutions.}
\end{figure}

The axion-dilaton-moduli solutions (figure~\ref{axidilmod})
correspond to a particle moving in $(\varphi,\beta)$ field space
which bounces off a potential $V(\varphi)\propto e^{-2\varphi}$:
\begin{eqnarray}
\varphi &=& \varphi_* - \ln2 + \ln
 \left( (\eta/\eta_*)^{\sqrt{3}(\cos\xi)} + (\eta/\eta_*)^{-\sqrt{3}(\cos\xi)} \right)
 \nonumber \\
\beta &=& \beta_* + \sqrt{3} (\sin\xi) \, \ln(\eta/\eta_*)
\nonumber
\end{eqnarray}

Recently \cite{DamHen00} it has been shown that this can lead to a
BKL-like chaos approaching the singularity in homogeneous
superstring cosmologies. In higher dimensional cosmologies the
antisymmetric field strength(s) of the superstring effective
actions have many more degrees of freedom each of which can act as
a wall in the dilaton-moduli phase-space. Damour and Henneaux
\cite{DamHen00} have shown that in a generic solution, i.e., where
all the axion-type degrees of freedom are non-vanishing then the
solutions are trapped in a compact region of phase-space, bouncing
off the walls infinitely many times in the approach to the
singularity. Of course, just as in the analogous case of Bianchi
IX cosmologies in general relativity, the low-energy effective
description itself breaks down before one reaches the singularity,
but this behaviour must be an important consideration for any
discussion of the nature, or avoidance, of cosmological
singularities in string theory.

\section{The pre big bang}

The most concerted campaign to address the practical implications
for cosmology of string theory has been the work of Gasperini and
Veneziano and their collaborators \cite{pbbwww} in developing the
{\em pre big bang} scenario.

There are three basic ingredients/assumptions:
\begin{enumerate}
\item The universe begins in the asymptotic past in a low-energy,
  weakly-coupled, vacuum state well-described by a low-energy
  effective supergravity theory.
\item Small fluctuations about this vacuum state lead to gravitational
  instability~\cite{collapse} and run-away evolution (collapse in the
  conformally related Einstein frame~\cite{deflation}) towards a
  high-curvature, strongly-coupled regime.
\item This run-away instability must be tamed by stringy effects and
  transformed into a conventional expanding cosmology, possibly by
  T-duality which could map a high-curvature regime to a low-curvature
  one.
\end{enumerate}

The pre big bang is an honest attempt to use the tools available
to us from string theory to make cosmological predictions. It has
incurred the wrath of those who seek a dynamical explanation of
the peculiar spatial homogeneity and flatness of the standard hot big bang
model (such as that offered by inflation~\cite{arl}) because the initial
state must be at least close to the homogeneity and flatness
sought~\cite{tuning}.
%
But if a model such as the pre big bang could been shown
to be in better agreement with observational data than, say,
inflation, then we would have to learn to live with the initial
conditions that required.
%
A more pragmatic objection is that the well-understood low-energy
effective solutions still contain a singularity at the end of the
pre big bang phase and this can only be avoided by introducing
much less-well understood loop- or string-corrections as the
strong-coupling or high-curvature regime is reached~\cite{exit}.

The pre big bang does offer possible observational tests. The pre
big bang phase has a causal structure similar to that of
conventional inflation and can lead to the generation of
effectively classical fluctuations on super-Hubble scales from an
initial vacuum state. In the pre big bang the spectrum of metric
perturbations is sharply peaked towards the smallest scales (those
leaving the horizon close to the singularity). This could give a
stochastic background of gravitational waves with a steep blue
spectrum \cite{BGGV} (quite different to the almost scale-invariant spectra
from inflation). But it would leave effectively no density
perturbations on large scales at the start of the hot big
bang~\cite{BGGMV}.

Axion-type fields would have to play a crucial role if the pre big
bang (rather than inflation) is to provide a mechanism to seed the
large-scale structure in the universe. This is because the kinetic
coupling of the axion field can produce an almost scale-invariant
spectrum of fluctuations~\cite{CEW97}. However because the background
axion density vanishes asymptotically, these are {\em isocurvature}
fluctuations. Durrer and collaborators have estimated the power
spectrum of the cosmic microwave background anisotropies that might
result from this novel isocurvature source if the axion remained
decoupled~\cite{seeds}. While another distinctive prediction, the
latest observations of the cosmic microwave background anisotropies
may be hard to reconcile with this model \cite{Ruth}.

Instead one may be forced to consider a pre big bang model that
mimics the {\em adiabatic} primordial perturbation spectra produced by
inflation~\cite{arl}. It is not widely recognised that in the presence of
non-adiabatic perturbations, the large-scale curvature
perturbation produced, say, during a pre big bang phase provides
only a lower limit on the large-scale curvature perturbation at
later times. Isocurvature modes provide an additional source for
curvature at later times on arbitrarily large scales if they are
coupled to the dominant energy density, as has recently been explicitly
demonstrated in the case of two-field inflation~\cite{GWBM}. It is
quite possible that isocurvature axion fluctuations
could be converted into adiabatic curvature perturbations either
at, or after, the transition to the post-big bang
evolution.\footnote{Since this talk was given two papers
\cite{EnqvistSloth,LythWands} have given explicit demonstration of
this effect due to late-decay of a massive `curvaton' field. A
distinctive observational signature in such a model would be a
direct correlation between the curvature perturbation and any
residual isocurvature perturbation as they would both be derived
from the same curvaton field fluctuations.}

\section{Brane-worlds}

All the string-inspired cosmological solutions I have discussed
thus far are based upon a traditional Kaluza-Klein treatment of
the extra-dimensions where the role of the higher-dimensional
geometry can be reduced to a collection of effective fields in a
more or less conventional four-dimensional theory. In the last
couple of years a new approach to dimensional reduction has
attracted much attention where matter fields may be constrained to
a lower-dimensional surface. This may be valid as a low-energy
description of some string theory solutions including branes,
where the end-points of strings describe gauge fields on the
lower-dimensional surface \cite{Polchinski}, or the Horava-Witten
model where gauge fields live on orbifold fixed planes
\cite{HW,LOSW}.

If only gravitational fields are able to explore the
higher-dimensional model then this raises the possibility that the
extra dimensions may be much larger than previously assumed, as the
four-dimensional nature of gravity is only tested on scales above
about a millimetre~\cite{ADD}. As discussed earlier, the higher
dimensional (fundamental) Planck scale may be much smaller than that
which is apparent in four-dimensions if the volume of the extra
dimensions is large compared with the fundamental Planck scale. This
offers a novel geometrical perspective on the hierarchy between
different energy scales (such as the disparity between the Planck and
suspersymmetry scales) found in particle physics, and intriguingly
suggests that the regime of quantum gravity may be closer than we
thought.

There have been many different approaches to studying the possible
cosmological consequences of the brane-world proposal. I will
follow my personal bias and restrict my discussion to just one
arising from a paper by Randall and Sundrum in 1999 which
allows us to study the role of the higher dimensional geometry at
the expense of the possible effects of additional bulk fields such as
the dilaton and form fields.

\subsection{Brane-world cosmology}

It was noted by Binetruy, Deffayet and Langlois~\cite{BDL} that
cosmological solutions based on a naive application of the
conventional 4D form of Einstein's equations could be incorrect even
at low energies in the brane-world scenario in 5D Einstein gravity.
They showed that an observer on a 3-brane with surface density
$\sigma$ embedded in five-dimensional Minkowski spacetime sees a
Hubble expansion rate $H$ given by the very unconventional Friedmann
expansion \cite{BDL}
\begin{equation}
\label{quadFriedmann}
3H^2 = \left( \kappa_5^2 \sigma \right)^2
\end{equation}

At around the same time Randall and Sundrum \cite{RSII} proposed a
brane-world based upon a flat (Minkowski) 3-brane embedded
in curved five-dimensional anti-de Sitter spacetime. This is a
consistent solution in five-dimensional general relativity
if the brane surface density $\sigma$ has a
critical value, $\lambda_{RS}$ (see below), which exactly cancels
the gravitational effect on the brane of the (negative)
cosmological constant in the bulk.
More remarkably Randall and Sundrum showed, by considering bulk
metric perturbations, the existence of a zero-mode solution in the
bulk which yields a massless 4D-graviton on the brane. The correct
Newtonian limit can then be recovered on length-scales greater
than either the size of the extra dimension, or the anti-de Sitter
curvature scale, whichever was the smaller.
Thus the four-dimensional Newtonian limit is recovered even when
the extra dimension is non-compact. The magnitude of the
apparent 4D Planck scale is again a derived quantity, in
this case proportional to the brane tension. Thus in this model
the brane tension must be positive (in contrast to another
proposal of Randall and Sundrum's where the visible brane tension was
taken to be negative~\cite{RSI}.)

\begin{figure}[t]
\centering
\includegraphics[scale=0.7]{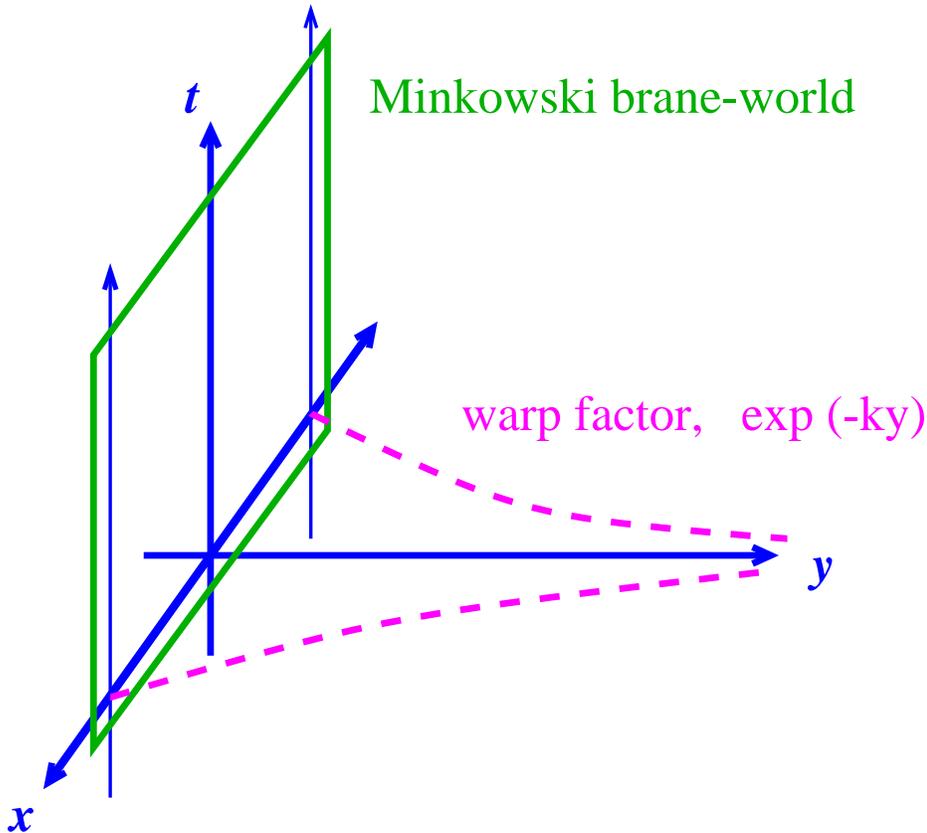}
\caption[AdSbrane]{\label{AdSbrane} Embedding of Randall-Sundrum
  (Minkowski) brane in five-dimensional anti-de Sitter spacetime in
  Poincar\'e coordinates: $ds^2=e^{-2k|y|}\eta_{\mu\nu}dx^\mu dx^\nu +
  dy^2$.}
\end{figure}

It should be emphasised that the Randall-Sundrum model falls into the
class of string-inspired rather than string-derived solutions.  It
draws on the notion of localised matter fields on a brane, but then
invokes Einstein gravity in a 5D vacuum. This might be contrasted, for
instance, with the earlier work of Lukas, Ovrut and Waldram
\cite{LOSW} who went to considerable lengths to derive a 5D model from
Horava-Witten theory which inevitably includes scalar fields and form
fields in the bulk. They found an analogous solution for Minkowski
3-branes embedded in a curved (but conformally flat) 5D spacetime.
However the simplicity of Randall and Sundrum's model has attracted
considerable attention.

In many ways studying the Randall-Sundrum model of brane-world gravity
is analogous to using Brans-Dicke gravity to model conventional string
theory solutions. The full theory is expected to have many more light
degrees of freedom in the perturbative limit whose couplings and
interactions could play an important role, but it is useful to have a
simplified but self-consistent model in which one can study the
non-linear gravitational dynamics. It is in this spirit that I will
discuss brane-world cosmologies based on the Randall-Sundrum model.

A negative bulk cosmological constant $\Lambda_5<0$ plus a
positive brane tension $\lambda>0$ is just what is required to
recover a viable Friedmann equation at low energies
($\rho\ll\lambda$) on the brane with total surface density
$\sigma=\lambda+\rho$ \cite{Csaki}. The modified Friedmann equation
(\ref{quadFriedmann}) becomes
\[
3H^2 = {1\over2}\Lambda_5 + \kappa_5^4 \left( \lambda + \rho \right)^2
\]
For a suitably chosen value of brane tension
$\lambda_{RS}^2=-\Lambda_5/2\kappa_5^4$ we have
\begin{equation}
 \label{modFriedmann}
 3H^2 = \kappa_4^2 \rho + \kappa_5^4 \rho^2
\end{equation}
where the four-dimensional Newton's constant is
$\kappa_4^2/8\pi=\kappa_5^4\lambda_{RS}/4\pi$ and we
recover the standard Hubble expansion rate for $\rho\ll\lambda_{RS}$.
Note that once again there may be a hierarchy between the four
dimensional Planck scale and the fundamental higher-dimensional
Planck scale if the brane tension $\sigma$ is significantly below
the Planck scale.

The brane-world thus offers a novel perspective on the cosmological
constant problem~\cite{self-tuning,Grojean}.  Instead of asking why
the vacuum energy density of our world is so small, the question
becomes one of why the brane surface density is so close to the
critical Randall-Sundrum value.

\begin{figure}[t]
\centering
\includegraphics[scale=0.7]{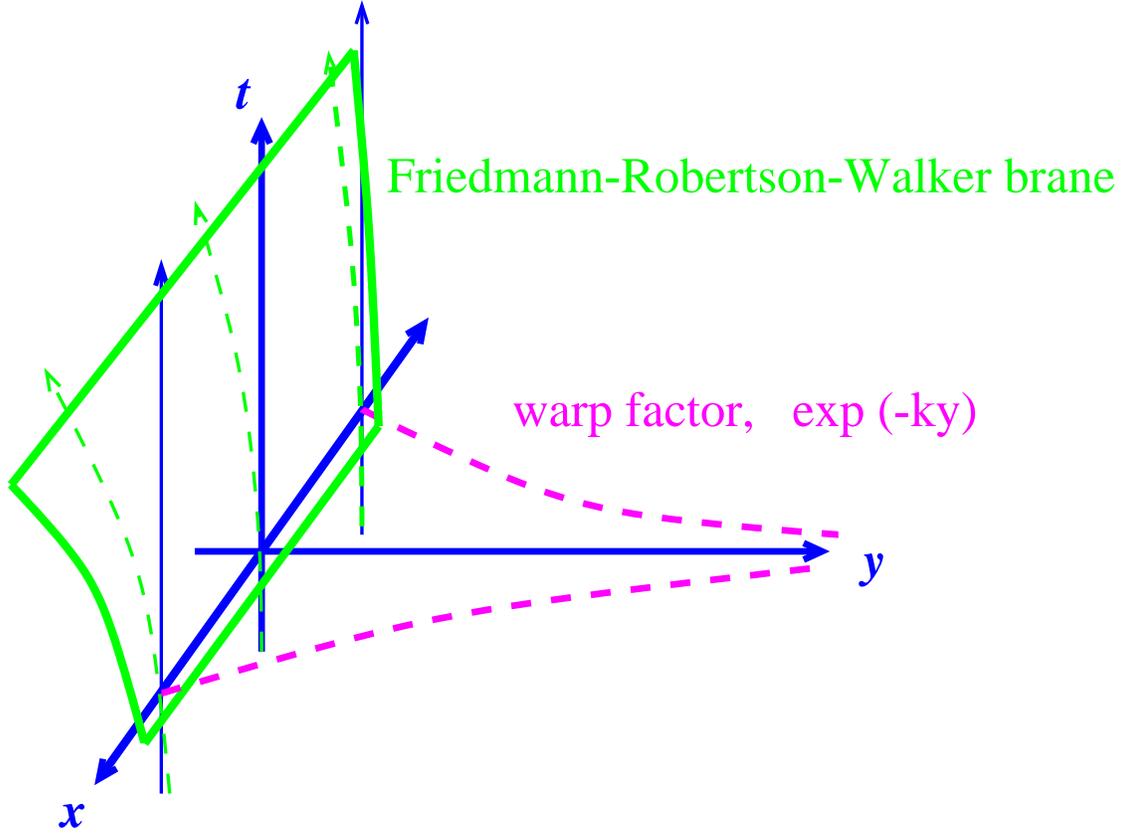}
\caption[FRWbrane]{\label{FRWbrane} Embedding of FRW brane in
  five-dimensional anti-de Sitter spacetime using same Poincar\'e
  coordinates in the bulk as Fig.~\ref{AdSbrane}.}
\end{figure}

The modified Friedmann equation admits a non-standard high-energy
regime for $\lambda_{RS}<\rho\ll M_5$ where one expects the
equation~(\ref{modFriedmann}) still to apply while matter fields
remain confined to the brane. In particular it would lead to a
modified evolution during inflation. The additional quadratic term
in the Friedmann equation leads to additional Hubble damping which
assists coventional slow-roll inflation driven by an inflaton
which is one of the matter fields on the brane~\cite{MWBH} and can
even yield inflation for potentials that would otherwise be too
steep~\cite{CLL,Sahni}.

Randall-Sundrum cosmology offers not just a modified system of
effective 4D dynamics but also a genuinely five-dimensional
perspective on our cosmological models. The maximal symmetry of the 3D
spatial slices of an FRW brane imposes a strong constraint on the
allowed 5D geometry. This enables one to find the explicit form of the
bulk solution for any non-static\footnote{The only exception I know of
  to this generalised Birkhoff theorem is the {\em static} FRW
  solution of Gergely and Maartens~\cite{GM}.}
FRW brane-world~\cite{BDEL} and this bulk solution turns out to be
Schwarzschild-anti-de Sitter spacetime \cite{MSM,BCG}.
After years of explaining to your mother (or your undergraduate
students) that it is not necessary to imagine that our universe is
expanding `into anything', it is now perfectly acceptable for you
to tell them that our universe could be expanding `into' a
five-dimensional Schwarzschild-anti-de Sitter bulk.

The Schwarzschild mass of the bulk spacetime turns up as an additional
constant of integration in the most general 4D modified Friedmann
equation, controlling the energy density of a dark radiation term
\cite{BDEL}. \footnote{One of the many intriguing aspects of the
  brane-world scenario is the holographic principle which asserts that
  a bulk gravitational field can be described by a non-gravitational
  field theory on the boundary of spacetime. The most famous example
  is the entropy of a black hole which is proportional to the area of
  its event horizon rather than its volume. Similarly the
  gravitational field of the bulk black hole turns up as an effective
  radiation density on the brane \cite{Gubser}.} This is an example of
a five-dimensional effect that (unlike the quadratic density term) is
not automatically suppressed at low energies. However if this dark
radiation is not to disrupt the standard successful model of
primordial nucleosynthesis its dynamical effect is constrained to
always be small.

\subsection{Covariant brane-world gravity}

The modified Friedmann equation is an example of non-linear
gravitational solutions that are possible in the Randall-Sundrum
model. An elegant way to describe the effective gravity seen by an
observer on the brane was presented by Shiromizu, Maeda and
Sasaki~\cite{SMS} who derived the effective Einstein equations for
an observer on a 3-brane embedded at a $Z_2$-symmetric fixed point
in a five-dimensional Einstein gravity.

Gauss's equation relates the intrinsic curvature of the
five-dimensional spacetime to the intrinsic curvature of the brane
and the extrinsic curvature of its embedding, $K_{\mu\nu}$. For a
brane located at a fixed point in a $Z_2$-symmetric bulk the
extrinsic curvature is completely determined by the tension $\lambda$
and energy-momentum tensor, $T_{\mu\nu}$, on the brane
\begin{equation}
K_{\mu\nu} = \kappa_5^2 \left( - T_{\mu\nu} + {1\over3} (T-\lambda)
  g_{\mu\nu} \right)
\end{equation}
The case of an asymmetric bulk is considered in Ref.\cite{Carter}.
The observer sees only the matter fields, with energy-momentum
tensor $T_{\mu\nu}$, the induced metric on the brane,
$g_{\mu\nu}$, and the projected five-dimensional Weyl tensor
$E_{\mu\nu}$ which appear in the equations for the 4D Einstein
tensor on the brane \cite{SMS}
\begin{equation}
\label{modEinstein}
G_{\mu\nu} + \Lambda_4 g_{\mu\nu} = \kappa_4^2 T_{\mu\nu} + \kappa_5^4
\Pi_{\mu\nu} - E_{\mu\nu} \,,
\end{equation}
where the effective cosmological constant on the brane is
\[
\Lambda_4 = {1\over2}\Lambda_5 + {\kappa_5^4\over12}\lambda^2 \,,
\]
and the quadratic energy-momentum tensor
\[
\Pi_{\mu\nu} \equiv -{1\over4}T_{\mu\lambda}T^\lambda_\nu + {1\over12}
TT_{\mu\nu} +{1\over8} g_{\mu\nu} \left(
  T_{\lambda\kappa}T^{\lambda\kappa} - {1\over3} T^2 \right) \,.
\]
represents the effect of high-energy corrections to the local
gravitational field, such as the $\rho^2$ term in the modified
Friedmann equation.

The effect of the non-local 5D gravitational field on the brane is
described by the projected Weyl tensor which appears as an
effective energy-momentum tensor
\[
\tilde{T}_{\mu\nu} \equiv -\kappa_4^{-2} E_{\mu\nu}
= \left({4\over3}\tilde{u}_\mu\tilde{u}_\nu + {1\over3}g_{\mu\nu}
\right) \tilde\rho + \tilde\pi_{\mu\nu} \,,
\]
with energy density $\tilde\rho$, momentum $\tilde\rho\tilde{u}^\mu$ and
anisotropic pressure $\tilde\pi_{\mu\nu}$. The evolution of the
Weyl energy and momentum on the brane follows from energy-momentum
conservation equations, driven by the Weyl pressure and
coupled to the quadratic tensor $\Pi_{\mu\nu}$~\cite{BMW2}.

The isotropic Weyl-pressure is $\tilde\rho/3$ due to the tracefree
property of the Weyl tensor, but the anisotropic pressure is in
general undetermined by the 4D equations and hence this system of 4D
equations in not closed. Hence it may be of limited use in actually
deriving brane-world solutions unless the anistropic stress from the
bulk is known or can be prescribed for some reason, as in the case of
FRW cosmologies where all anisotropies are required to vanish simply
by the symmetry of the spacetime. When other fields are included in
the bulk (as for instance in the heterotic M-theory solutions of
Lukas, Ovrut and Waldram~\cite{LOSW}) there are further undetermined
terms in the projected 4D equations on the brane representing energy
transfer to or from the bulk fields and this need not be vanishing
even for FRW cosmologies on the brane~\cite{MW}. Nonetheless the 4D
effective equations are required for the correct 4D interpretation of
the 5D solutions.

For example, it is often supposed that moving matter on the brane must
`lose energy' to the gravitational field in the bulk~\cite{Hebecker},
whereas in five-dimensional vacuum Einstein gravity, the local
energy-momentum conservation, $\nabla^\mu T_{\mu\nu}=0$, is guaranteed
by the Gauss-Codazzi equation~\cite{SMS}. The answer is that moving
matter on the brane {\em can} generate bulk gravitational waves but
that this is due to energy-momentum transfer between the quadratic
tensor, $\Pi_{\mu\nu}$, and the effective Weyl-fluid. The covariant
derivative of Eq.~(\ref{modEinstein}), plus the Bianchi identity
$\nabla^\mu G_{\mu\nu}=0$, yields
\[
\nabla^\mu E_{\mu\nu} = \kappa_5^4 \nabla^\mu \Pi_{\mu\nu} \,,
\]
without affecting the local conservation of the conventional
four-dimensional energy-momentum tensor $T_{\mu\nu}$.

\subsection{Cosmological perturbations}

Although quadratic terms in the modified Friedmann equation may be
significant in the high-energy regime ($\rho>\lambda_{RS}$) the
effect at low-energies (say, after nucleosynthesis) becomes
negligibly small. Any evidence of an unconventional past could
show up in any primordial perturbations generated at early times.

If we wish to consider arbitrary linear perturbations about an FRW
brane in a (Schwarzschild-)anti-de Sitter bulk, then the projected
field equations do not form a closed system due to the possibility
of anisotropic stresses.
The standard approach to studying linear perturbations in a higher
dimensional theory is to decompose arbitrary perturbations into
harmonics on the spatial sections. If the wave equation is
separable then one can follow the time-dependence of each spatial
mode separately. This is always possible, for instance, for
perturbations about an FRW background where 3D spatial
eigenmodes are Fourier modes (in flat space) or (hyper-)spherical
harmonics in curved space . However, for an FRW brane it is not in
general possible to separate the time-dependence of each 3D mode
from the bulk-dependence in a brane-based coordinate system. One
is then left with a partial differential equation to solve for
which it has proved difficult to come up with analytic results except in
special cases.

If the bulk is 5D anti-de Sitter it is always possible to write down
the general separable solution in a static coordinate system
\cite{DD,KS}. This indeed was the approach used to study the Newtonian
limit by Randall and Sundrum in their original paper.  But although a
Minkowski brane remains at fixed location, an FRW brane will be moving
with respect to the static bulk coordinates and applying the correct
boundary conditions at the brane couples together different bulk
modes. So although the {\em general} solution is known, the {\em
  particular} solution that obeys the correct boundary conditions is not.

As a consequence there is no zero-mode solution for the bulk
gravitational field in a simple FRW cosmology such as a radiation-
or matter-dominated universe. Although this mode mixing will be
suppressed on scales where the cosmological expansion/brane motion
can be neglected, the lack of quantitative results
means that the magnitude of any deviations from the standard
evolution of cosmological perturbations remains uncertain.
On small (sub-Hubble) scales we expect the Minkowski results to
remain valid. And on large scales there is a solution where the
metric perturbations remain constant when all gradient terms can
be neglected~\cite{LMSW}. But around horizon-crossing, where the
cosmological expansion cannot be neglected, there could be some
mode-mixing~\cite{GRS}.

There have been many attempts to extend to the brane-world either
metric-based or covariant approaches to cosmological perturbation
theory, reviewed in Ref.~\cite{BMW2}. These have, however, largely
been restricted to developing the required formalism and have stopped
short of providing quantitative results.

The only cases in which the 5D perturbations are properly
understood are those special cases in which the bulk equations are
separable in a brane-based coordinate system. For 5D vacuum
Einstein gravity this restricts the brane to only de Sitter (or
anti-de Sitter or Minkowski) cosmologies in 5D anti-de Sitter.
Although a mathematically special case, de Sitter provides an
idealised model for conventional slow-roll inflation. Thus one can
calculate the spectrum of large-scale metric perturbations
generated from initial vacuum fluctuations that cross outside the
cosmological horizon during the de Sitter
phase~\cite{MWBH,LMW,BMW1}. What one finds is that large-scale
perturbation spectra are {\em only} generated in the zero-mode,
i.e., the massless 4D graviton, and the massive bulk modes are
never excited. There is an enhancement to the amplitude of
gravitational waves (tensor modes) generated at a given energy
scale if that energy is greater than $\lambda_{RS}$. But if
inflation is driven by a scalar field on the brane then the scalar
curvature perturbations are also enhanced and enhanced more than
the gravitational waves~\cite{MWBH}. However, surprisingly, the
consistency relation found betweeen the ratio of tensor to scalar
amplitudes and the tilt of the gravitational wave spectrum in
conventional single-field inflation still holds~\cite{Huey}, and
the Randall-Sundrum gravity misses another opportunity to
distinguish itself observationally from a purely four-dimensional
theory~\cite{LidTay}.

\section{Conclusions}

String cosmology is far from being a mature subject in which there
is any great consensus about how to tackle cosmological questions
or even which cosmological questions should be tackled. In this
review I have naturally focussed on those questions that interest
me and the solutions I have sought. My own hope is that it is
through the nature of primordial perturbation spectra being
measured with ever-increasing precision by observational
cosmologists that we may learn about high-energy phenomena in our
Universe.

String theory requires us to contemplate cosmological solutions in a
higher-dimensional spacetime. In traditional Kaluza-Klein
compactifications this yields a lower-dimensional spacetime described
by dilaton-gravity theory with additional low-energy matter fields
arising from the `internal' geometry of the extra dimensions.
Higher-dimensional effects, corresponding to excitations of the higher
harmonics on the internal space, are then supressed at energy scales
below that set by the compactification scale.

String theory is best understood in terms of effective field
theories in the low-energy limit. The pre big bang scenario is an
honest attempt to use the low energy effective actions to describe
a cosmological prehistory of our Universe. To date it has failed
to provide observational predictions that are in any better
agreement with the cosmological observations than those of
conventional inflation models. At the same time questions remain
about the nature of the transition from pre- to post-big bang
phase where a deeper understanding of string (or M-) theory may be
required.

A different model for our place in the higher-dimensional
spacetime has been provided by the brane-world scenario in which
matter fields are restricted to lower-dimensional surfaces, or
branes. In order to study the consequences of this novel
geometrical setting I have focussed on the simple model proposed
by Randall and Sundrum of a single brane embedded in 5D vacuum
spacetime described by Einstein gravity. This is simple enough for
the homogeneous and isotropic cosmological solutions to have been
well-understood, but the understanding of cosmological
perturbations is still largely restricted to de Sitter inflation
and the large-scale limit. There is some hope that novel 5D
effects may survive even at low energies due to the non-local
gravitational field but these remain the hardest to calculate and
their effect remains uncertain.

If string theory is to be vindicated experimentally then it seems
likely that cosmology will provide the key. But the challenge
remains to provide observational predictions that, without
violating existing experimental bounds, go beyond those effects
already known to the four-dimensional cosmologist.

\ack
I am grateful to the meeting organisers for their kind hospitality
during my stay in Cape Town.  My work is supported by the Royal
Society.






\end{document}